# Triple Junction and Grain Boundary Influences on Climate Signals in Polar Ice

Thomas M. Beers[1,2], Sharon B. Sneed[1], Paul Andrew Mayewski[1,2], Andrei V. Kurbatov[1,2] and Michael J. Handley[1]

1. Climate Change Institute, University of Maine.
2. School of Earth and Climate Sciences, University of Maine.

**Abstract:** The Climate Change Institute's W. M. Keck Laser Ice Facility laser ablation inductively coupled plasma spectrometer (LA-ICP-MS) yields a sample every 121 micrometers, a resolution on the scale of ice crystal triple junctions and grain boundaries in ice cores. Recent publications suggest that these features can allow amplification of impurity concentrations, and allow migration through veins potentially obscuring climate signals preserved in polar ice. LA-ICP-MS data reveal that such features modify these signals by less than 6% in the case of Na, Ca, and Fe based on the examination of GISP2 ice deposited at the beginning of the Holocene.

**Introduction:**

The laser ablation inductively coupled plasma mass spectrometer (LA-ICP-MS) at the Climate Change Institute provides ~10,000 samples per meter, a resolution that potentially allows researchers to see storm signals in low snow accumulation areas, and in highly compacted areas, such as basal ice. Following pioneering LA-ICP-MS ice sampling by Reinhardt et al., 2001 & 2003, Sneed et al., 2015 developed a Sayre Cell that holds and keeps frozen meter long sections of ice while being ablated (physically blasted from the ice surface) for sampling by a New Wave Laser (213 nm in phase wavelength). Laser spot size and x-axis velocity (the laser is moved automatically along the core surface) produce a sampling resolution of 121 micrometers (μm) (figure 1).

Triple junctions and grain boundaries (here on collectively referred to as TGs) have been observed to hold higher concentrations of impurities specifically sulfur compounds in several studies (Mulvaney et al., 1988, Wolff et al., 1988, Faria et al., 2010, Rempel et al., 2001, Rempel et al., 2002, Rempel et al., 2003, Barnes et al., 2003, Cullen et al. 2006, Durand et al., 2006, Obbard et al., 2003). The potential increased concentration in TGs relative to surrounding grains has the potential to obscure climate signal preservation. If climate signals are not preserved within grains as they are between grains, grain size would dictate the maximum sampling resolution (on the order of millimeter scale) potential for ice cores. We investigate the role TGs play in the preservation of climate signals, specifically sodium (Na), calcium (Ca), and iron (Fe), in attempts to quantify any potential disruption of these records. We chose Na, Ca, and Fe for this study as each has demonstrated proxy value in ice core climate reconstruction. In the case of the GISP2 ice core (central Greenland), Na is a soluble marine sea salt species primarily transported by the Icelandic Low (Mayewski et al., 1997), Ca is terrestrial dust linked with the northern hemisphere westerlies and the Siberian High (Meeker and Mayewski, 2002), and Fe is sourced from iron rich minerals dissolved in the atmosphere (Dedick et al., 1992). All three elements are proven to be reliably measured using the LA-ICP-MS system (Sneed et al., 2015), used directly (Sneed et al., 2015, Mayewski et al., 2013) and indirectly (Meese et al., 1997) for dating, and offer a test for insoluble and soluble species post-depositional movement.

**Methods:**

For this study we chose to laser ablate Greenland Ice Sheet Project Two (GISP2) ice from a depth of 1677-1678 meters (11,628 – 11,652 GISP2 a BP, Meese et al., 1997). We chose this depth range because the average grain size undergoes a drastic increase at this depth from 2.3 to 4.6 mm coincident with the transition from the Younger Dryas cold event into the Holocene, respectively, and little flow disruption is recorded at this depth (Gow et al., 1997). The annual layer thickness only 25 m deeper at 1703 m is ~2.5 cm/yr as recorded by Alley et al., 1997, and Mayewski et al., 2013 find the same annual layer thicknesses at 1677m. We assume that sub-annual layers strike perpendicular to the long axis of the core on the millimeter scale as confirmed by several parallel ablation runs perpendicular to the core (demonstrating homogeneous concentrations within layers), and thin section studies by Gow et al., 1997 also find layers strike perpendicular to the long axis of the core. Concentrations of Na, Ca, and Fe are analyzed in several different runs throughout the meter. Only one element is measured per ablation run, thus three parallel runs are conducted beside each other (offset by less than 50 micrometers) to measure Na, Ca, and Fe.

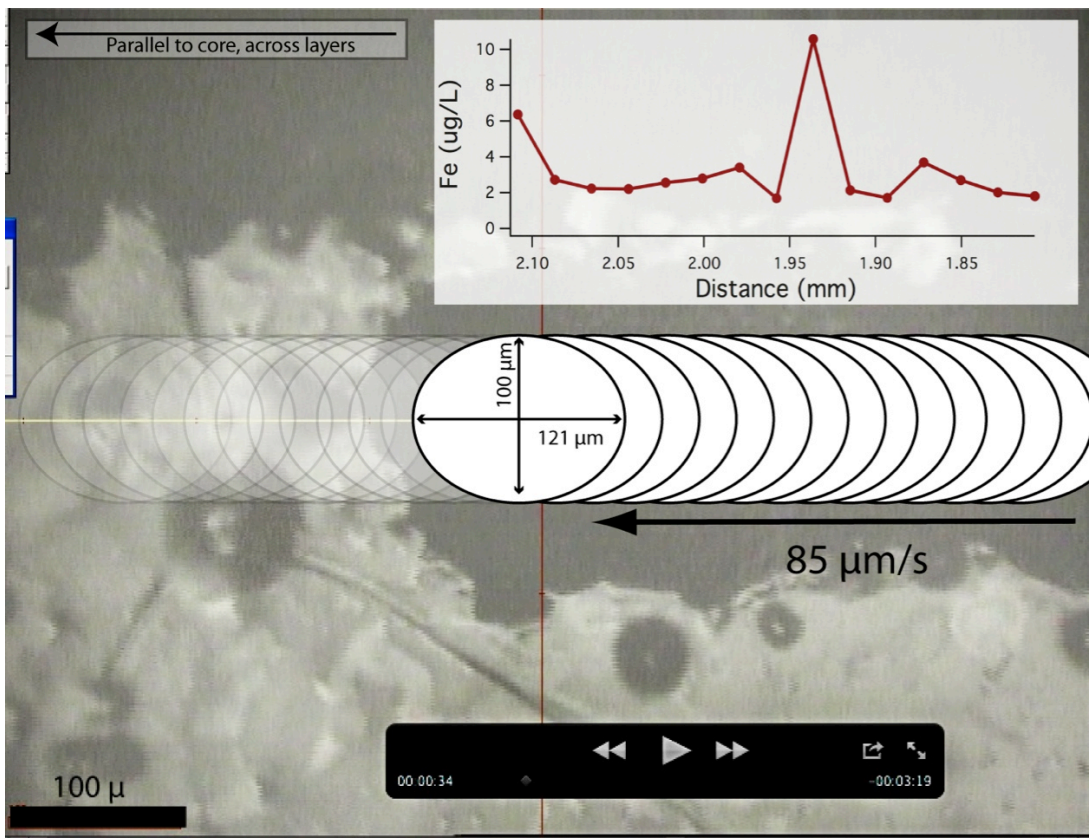

Figure 1: An example of ablations perpendicular to annual and sub annual layers. The laser ablation trench (direction represented by the arrow) just before it ablates the triple junction. The distance at which this triple junction is ablated is recorded, and then plotted against concentration. The sample resolution is represented in the trench by white ellipses (100 µm by 120 µm) and the analyzed concentrations for each sample plotted above. Notice the scale of the triple junction relative to the sample resolution. 6 samples record the triple junction concentration.

We use two different ablation run schemes to investigate if TGs concentrate impurities: (1) ablations perpendicular to layers, and (2) intra-feature ablation.

Ablation perpendicular to layers:

Two-centimeter ablation runs were preformed parallel to the long axis of the core, perpendicular to layers, and recorded with a microscope fitted camera (figure 1). Real time video allows identification of TGs as the ice is laser ablated, depths are recorded, and later plotted against concentration. Ablation runs are only 2 cm in length because of the limiting size of the ablation chamber, hence subsequent runs start precisely where the previous run ended and continue along the same strike as the previous run until a meter is completely and continuously analyzed. Figure 1 demonstrates the lasing process. Laser spot size is 100 µm and the laser ablates along the x-axis 21 µm per sample. This yields overlapping ellipses (100 µm by 121 µm) thus a feature such as a triple junction or grain boundary (~50 µm) will contribute to 42% of the concentration analyzed for 6 samples. As a result, higher concentrations expected in TGs relative to grains should be apparent in 6 samples. The (expected) higher concentrations in TGs are diluted by surrounding (expected) lower concentrations of intra-grain ice.

Ablations perpendicular to layers are in counts per second for Na, Ca and Fe and are converted into concentration using a transfer function calculated by ablating a frozen standard (see Sneed et al., 2015). TG positions in the ablation run are plotted against concentrations of Na, Ca, and Fe to investigate coincidence of variability in concentration with TGs over several years (24 years recorded by Meese et al., 1997) and with grains (average grain size 2.3 mm, Gow et al., 1997). This method does not allow us to decipher whether TGs result in, produce or are simply coincident with impurity concentration fluctuations, but it does provide frequency of coincidence.

Intra-feature ablation:

We performed intra-feature ablations parallel to layers to clarify if TGs actually are preserving higher concentrations than found inside grains, or if they are merely coincident with peaks observed in runs perpendicular to layers. In picking ablation runs parallel to layers (intra-layer) we attempted to avoid the concentration fluctuations that naturally occur from layer to layer. Ablating just one layer is more homogeneous in concentration than ablating multiple layers, as different seasons in Greenland are dominated by variations in atmospheric circulation, which strongly impacts fluctuations in concentration over seasons (Meeker and Mayewski, 2002). The more homogeneous concentration observed in a single season (layer, as compared to multiple layers) ideally emphasizes the role TGs play in preservation of Na, Ca, and Fe as we are comparing just TGs and grains, and excluding seasonal fluctuations.

Within the layers studied we start by ablating a grain boundary, through a triple junction, and end in a grain as the control in our sampling (figure 2). This operation of ablating just features increases the time (and distance) a specific feature is ablated, thus isolating TG concentrations from grain concentrations. As a consequence there is no question whether a peak just coincides with such a feature, rather than being caused by the feature. Causation would not be as easily investigated in ablations perpendicular to layers section because multiple layers are crossed, thus concentration fluctuations could not be solely attributed to TGs. The influences of TGs become apparent as we observe changes in background levels, and peak magnitudes when comparing triple junctions, grain boundaries, and grains within a single layer.

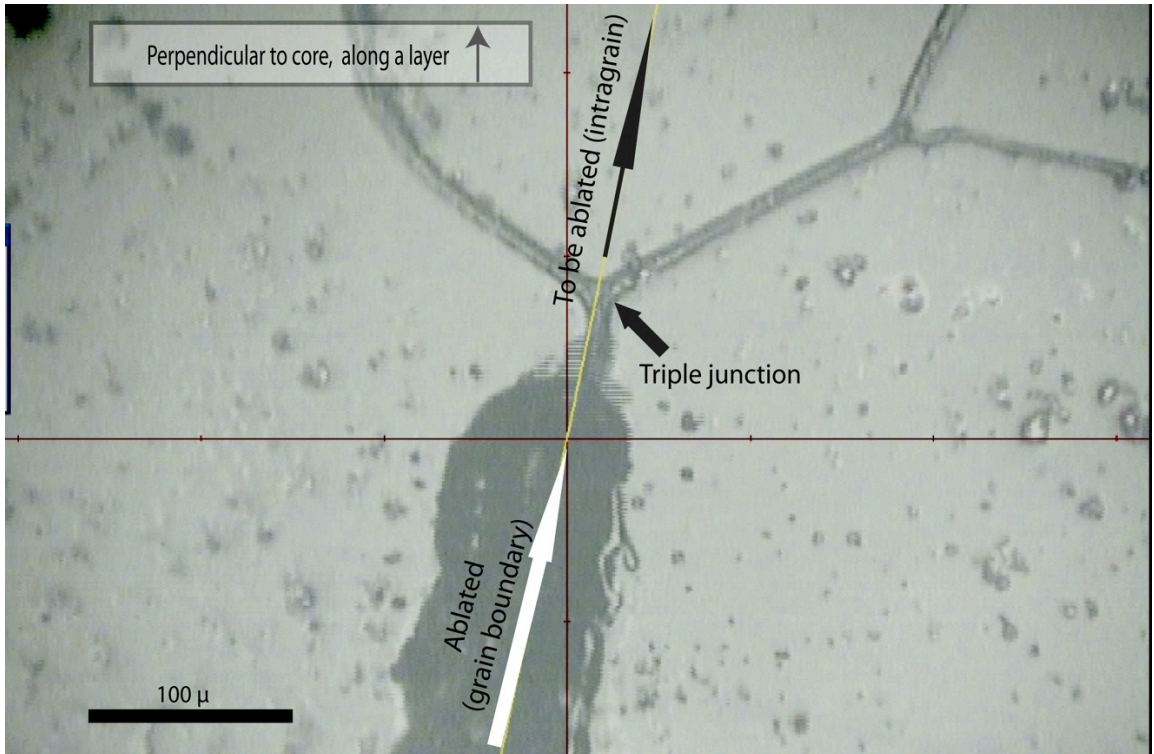

Figure 2: An example of an intra-feature ablation run, a laser ablation path through a grain boundary into a triple junction, ending in a grain. The lasing paths where chosen so that the triple junction would be at the center of run for ease of plotting locations of grain boundaries, triple junctions and intra-grain against concentrations.

**Results:**

Ablation Perpendicular to Layers:

For runs parallel to the long axis of the core concentrations of Na, Ca, and Fe vary greatly from annual layer to annual layer throughout the meter as seen in figure 3 - a plot of the first 9 cm of meter 1677. We focus on 2 cm of the core (1677.02 - 1677.04 m) as this section has a higher average concentration (Na: 8 µg/L, Ca: 133 µg/L, and Fe: 17 µg/L) for Ca and Fe than surrounding layers. The higher average concentration is of specific interest since we assume that the TG influence on concentrations could likely be emphasized within these layers, if it has an impact. The following runs across multiple layers (figure 4) are only 3 of 75 ablations preformed and observed, and are chosen for this paper as they demonstrate both grain boundaries and triple junctions in one 2 cm ablation run. Of the n=75 2 cm ablations, the 3 ablation runs in figure 4 demonstrate above average number of TGs ablated (>1 triple junction and >4 grain boundaries). Thus we are investigating the highest density of TGs observed in this study to be able to estimate a maximum percentage of climate signal disturbance in GISP2 at the beginning of the Holocene.

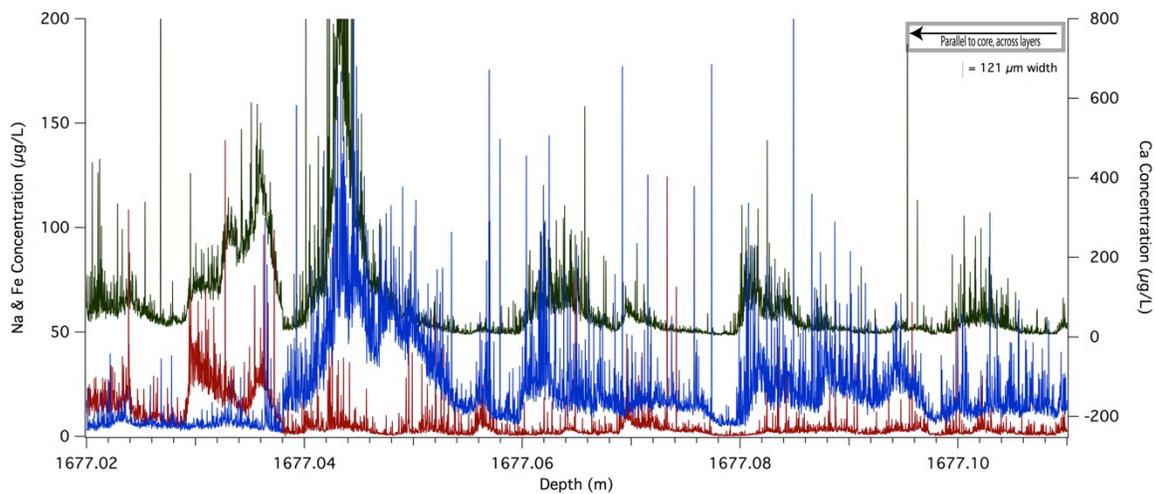

Figure 3: Concentrations of Na (blue), Ca (green), and Fe (red) analyzed from 1677.02 to 1677.11 m in GISP2 ice. During this study we preferentially ablated layers of higher concentration (1677.02-1677.04 m) during intra-feature ablations, as we assume these layers emphasize triple junction and grain boundary influences on concentration. The sample resolution is denoted by the width of the thin vertical line (top right).

Plotting TGs against concentration over 2 cm sections reveals several peaks within TGs and grains (figure 4), not just in TGs. The vertical markers show where the lasing path crossed a TG and all areas on the plots not marked by a vertical marker are ice inside grains. For the Na run (figure 4A), 2 triple junctions and 7 grain boundaries were ablated. There are several Na peaks throughout the run, intra-feature and inside grains, and the highest background level of the run is inside grains, thus grains contain the highest concentration. Several one point peaks coincide with both triple junctions and grain boundaries, but none are above the peaks recorded inside grains. Similarly, the Ca lasing run (figure 4B) crosses 2 triple junctions and 5 grain boundaries, and peaks in concentration are found throughout. The highest concentrations analyzed reside inside grains and peaks coincident with TGs only occur in the first 6 mm of the run. Some of the lowest concentrations analyzed are coincident with both triple junctions and 2 grain boundaries from 9-18 mm of the run. The Fe ablation (figure 4C) crosses 3 triple junctions, and 4 grain boundaries, ablating within two grain boundaries that are along the strike of the run. Once again, the highest peaks and background levels are found inside grains, and features coincide with one-point peaks that are not above intra-grain peak concentrations. We did not calculate means and standard deviations to compare and contrast TGs and grains for this section of the study as we ablated across several layers (heterogeneous concentrations). If calculated, the means and standard deviations would be representative of seasonal fluctuations in chemistry and TGs influence, therefore confounding the intended investigation. A comparison of TGs and grains is discussed in the following intra-feature ablation section.

A. Na Ablation

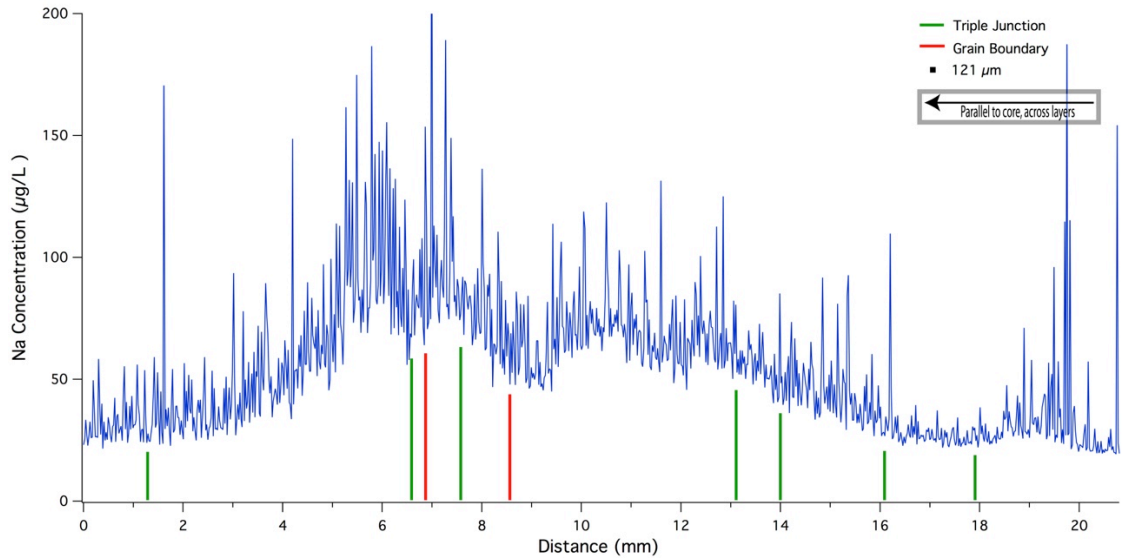

B. Ca Ablation

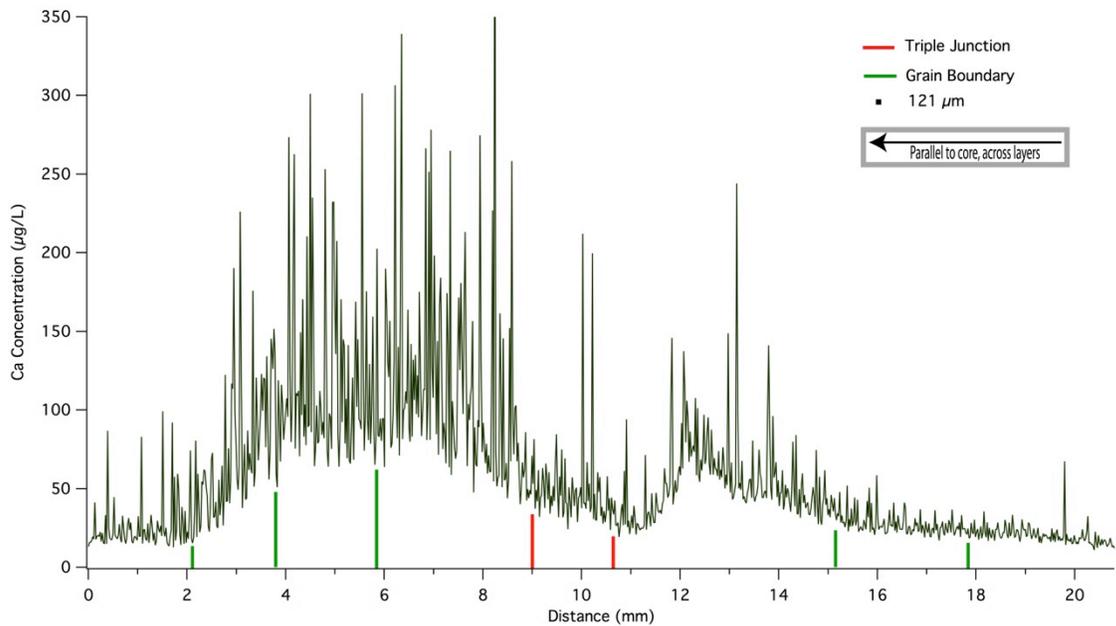

C. Fe Ablation

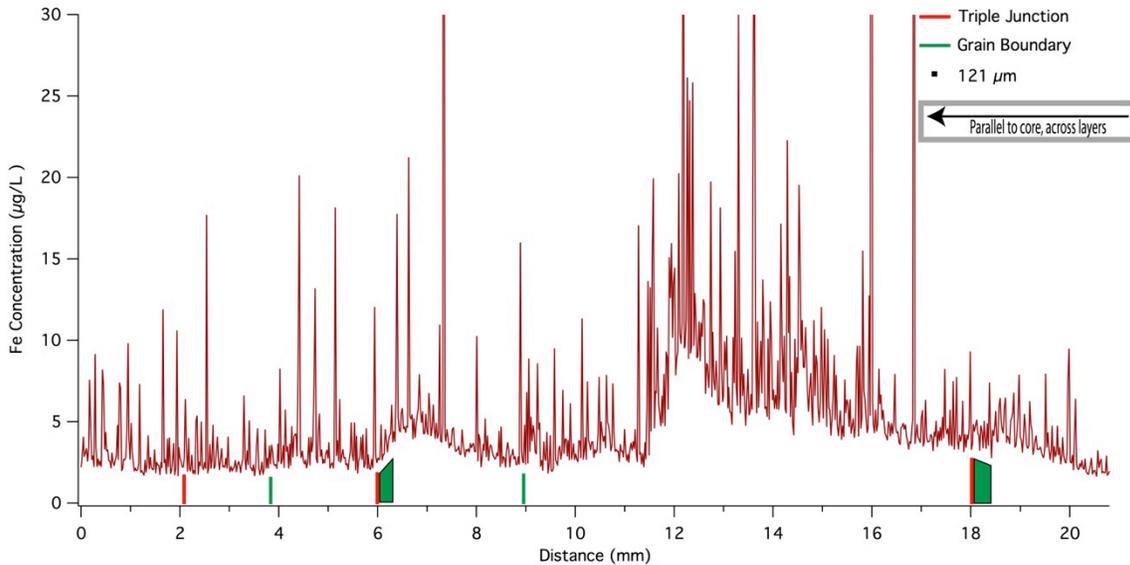

Figure 4: 2 cm ablation runs (A. Na, B. Ca, C. Fe) perpendicular to sub-annual layers of Na, Ca, and Fe. Vertical red lines below concentration plots denote locations of triple junctions, the thickness of which is representative of the scale of the feature. Similarly, green lines below concentration plots represent locations of grain boundaries, the thickness of which is representative of the scale of the feature. Green polygons represent ablation along a grain boundary (also to scale), thus not just crossing a grain boundary but ablating thru it. The sample resolution (121 µm) scale is shown in the upper right, notice that features are recorded in 6 samples for a feature ~50 µm in width.

Intra-feature ablation:

We preformed four lasing runs through a grain boundary into a triple junction, each ending intra grain to measure Na, Ca, and Fe concentrations throughout the TG and grain (see figure 2). These lasing runs appear in figure 5, plotting concentrations of an ablated grain boundary, triple junction, and inside grain ablation. Means and standard deviations for all runs of Na, Ca, and Fe are in table 1. Using the average of all four runs, grain boundary (GB) concentrations are Na (GBs = 68.1-+24.8 compared to inside the grain = 54.4-+10.6 µg/L), Ca (GBs =75.3-+63.4 compared to inside the grain = 45.3-+18.6 µg/L), and Fe (GBs = 2.9-+2.1 compared to inside the grain = 2.9-+1.8 µg/L). We find no difference between grain boundary and inside grain concentrations for Fe. Triple junction concentrations are not averaged as the sample size only accounts for <6 out of ~280 samples per run, which would produce a statistically insignificant number.

A. Na Intra-feature ablations

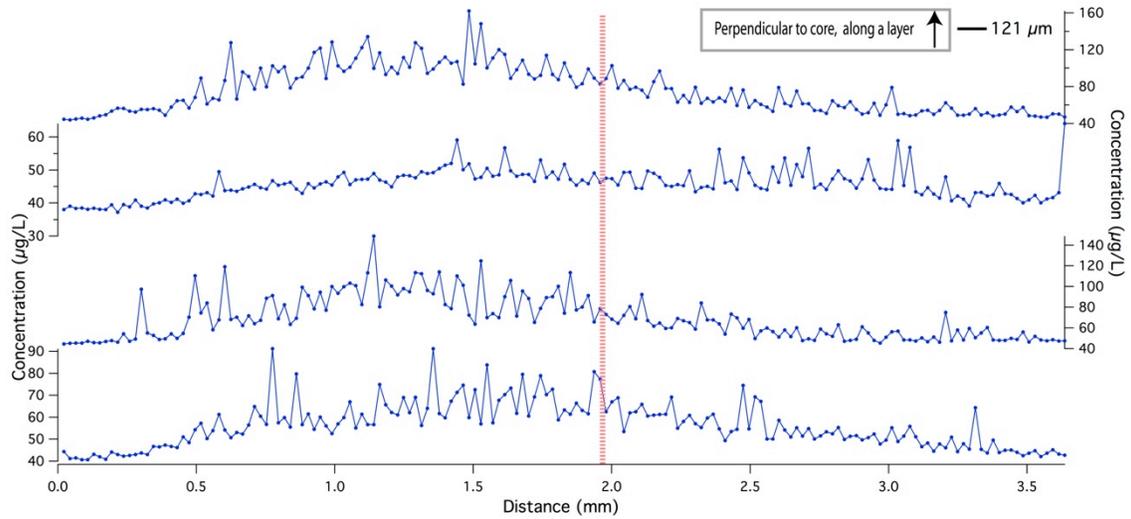
B. Ca Intra-feature Ablations
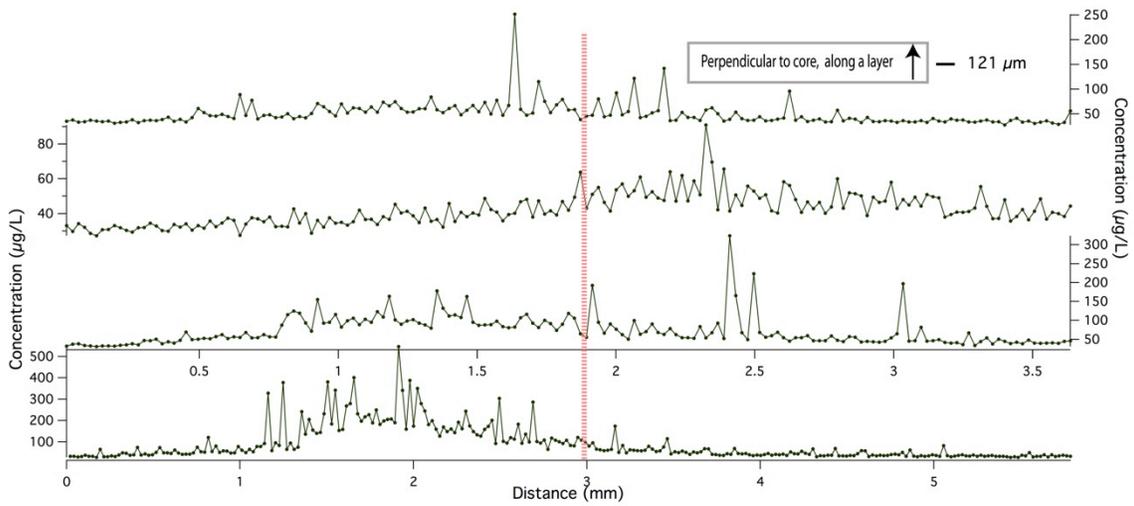
C. Fe Intra-feature Ablations
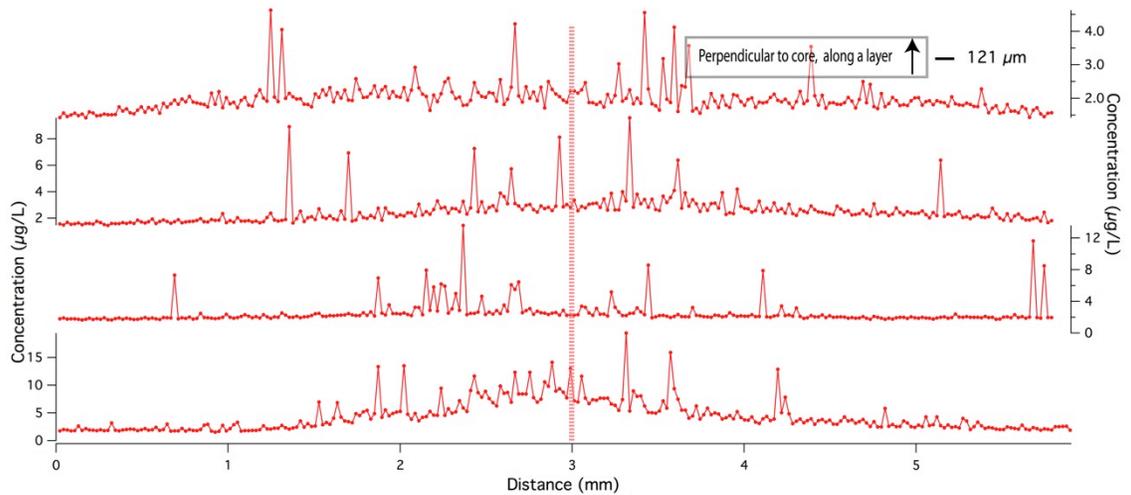

Figure 5. Intra-feature ablation runs (A. Na, B. Ca, C. Fe) perpendicular to the long axis of the core, intra-layer. The beginning of the run starts inside a grain boundary (left of the vertical marker), then ablates through a triple junction (dashed vertical red line, the width is to scale of the triple junction), ending inside a grain (right of the vertical marker). Four runs for each element were measured and triple junctions are centered in the run.

Table 1. Values of average concentrations and standard deviations over four runs, per element (plotted in figure 5). This table provides an estimate of concentration differences between grain boundaries (GB) and intra-grain spaces (INT).

|              | Na (µg/L) | Ca (µg/L) | Fe (µg/L) |
|---|---|---|---|
| Mean GB      | 68.1 | 75.3 | 2.9 |
| Std dev GB   | 24.8 | 63.4 | 2.1 |
| Mean INT     | 54.4 | 45.3 | 2.9 |
| Std dev INT  | 10.6 | 18.6 | 1.8 |

**Discussion:**

Grain boundaries contain higher Na and Ca concentrations than inside grain spaces based on intra-feature ablations within a single layer (be it sub-seasonal or annual), and no statistical difference is found for Fe. Ablations performed perpendicular to layers reveal that the higher concentrations found in grain boundaries do not contain peaks or background levels greater than surrounding inside grain peaks and levels when ablating across multiple layers, thus grain boundaries can concentrate impurities inside a layer but not above the surrounding annual/sub-annual layers' magnitude. Furthermore, we find that the spacing of predominant inside grain peaks is ~2cm (figure 3) comparable to annual layer thicknesses previously recorded by Alley et al., 1997 near this depth (~2.5 cm/yr at 1703 m) and Mayewski et al, 2013 at 1677 m.

Triple junctions, although only accounting for <6 out of ~280 samples per intra-feature run (and thus not statistically significant), do not coincide with anomalous peaks greater than surrounding spaces. If such anomalous concentrations reside in triple junctions there would be a relatively large peak in every intra-feature ablation run centered on the triple junction. The only ablation sampling in this study suggesting that triple junctions concentrate impurities is seen in figure 5C. Relatively higher concentrations of Fe (8 µg/L compared to 2 µg/L) are centered around a triple junction, but there is no difference in concentration between intra-granular and grain boundary concentrations.

In reviewing Durand, et al. 2006, we find an explanation for some of our observations. Soluble impurities, such as the majority of Na and a portion of Ca in GISP2, are 'dragged' by grain boundaries as grains grow. Insoluble particles such as dust (a large component of the Fe budget) as well as clathrates and bubbles act as pinning points. While ablating intra-feature within a single layer we observe that grain boundaries (GBs) do hold higher concentrations of Na (GBs = 68.1-+24.8 compared to inside the grain = 54.4-+10.6 µg/L) and Ca (GBs = 75.3-+63.4 compared to inside the grain = 45.3-+18.6 µg/L), but find no difference for Fe (GBs = 2.9-+2.1 compared to inside the grain = 2.9-+1.8 µg/L). Durand et al.'s work does not clarify our finding that the highest background levels and peaks analyzed are inside grains when ablating multiple layers. Our results suggest limited migration of Na, Ca, and Fe, as we sample with a 121 µm depth resolution, and cross <10 TGs that are ~50 µm in width, producing 60/1000 samples attributed to TGs. Thus, TGs account for less than 6% of samples at a 121µm sampling resolution.

Insolubles (in our study Fe - figure 5C) and clathrates (observed in this study as bubbles) may have acted to pin grain boundaries and triple junctions as suggested by Durand et al. (2006) causing hindered growth of grains. This hindered growth rate would then limit the dragging of Na and Ca in grain boundaries, displacing impurities to a lesser extent than if pinning were not occurring. Our work suggests that limited concentration or dragging of soluble Ca and Na occurs in grain boundaries during grain growth, but it also demonstrates that the vast majority (94%) of the Na and Ca can be found in-situ inside grains. Fe is found throughout TGs and grains with no statistical difference in concentrations.

Rempel et al., 2001 suggest that anomalous diffusion of impurities occurs with amplitudes of impurities preserved (specifically $H_2SO_4$) through veins (connected triple junctions and grain boundaries) when compared to surrounding ice velocities. Thus the climate signal could be displaced by up to 50 cm relative to the ice the impurities were originally deposited within. Rempel et al., 2001 cite ice matrix space problems (Wolff et al., 1996) and eutectic conditions (ex. $H_2SO_4$ solidus at -73° C) as the cause for the diffusion. Our study, albeit focused on one core section, finds sufficient space within the ice matrix for Na, Ca, and Fe, as we find all elements throughout grains and TGs with the highest concentrations found inside grains. Veins pose an interesting explanation for the concentration distribution of Na, and Ca in grain boundaries. If such species are in acid form the solidus is reached at a much lower temperature than existed in the GISP2 core at this depth (-32° C, Gow et al., 1997). Nonetheless, our study still finds the largest peaks undisturbed inside grains (not in veins), with less than 6% of Na, Ca, and Fe concentrations disrupted by grain boundary concentration and/or anomalous diffusion. If veins are conduits for Na and Ca, they are not the dominant ice matrix spaces Na and Ca reside in GISP2 ice used in this study.

Our work supports findings by Barnes et al., 2003 and Obbard et al., 2003 that suggest grain boundaries concentrate impurities, and triple junctions act only as a secondary location for impurities to reside when a concentration threshold is exceeded (when grain boundaries are occupied to saturation). Both studies also find Na and Cl in the crystal lattice, but attribute their location to dislocations in the lattice and isolated inclusions. Our findings additionally reveal Na, Ca, and Fe throughout the lattice, grain boundaries and triple junctions, thus we find these elements do not reside just in isolated inclusions, but are likely continuous dislocations within the matrix.

**Conclusions:**

This study demonstrates that features such as grain boundaries and triple junctions can concentrate impurities, but not above the magnitude of the surrounding annual layer signal peaks that are detected intra-granularly. Moreover, peaks in impurity concentration found in such features account for less than 6% of the total number of samples analyzed leaving more than 94% of the climate signal preserved at the micrometer scale in at least the GISP2 ice section sampled at 1677 m. These findings make the detection of potential storm signals in ice cores possible at the micrometer scale.

**Acknowledgements:**

We thank the W. M. Keck Foundation for funding the setup and refinement of the LA-ICP-MS system, the Advanced Manufacturing Center at the University of Maine for assembly of the LA-ICP-MS Sayre cell, Dr. Nicky Spaulding for helpful suggestions, and the National Science Foundation (grants 1203640, ANT-1042883) for support of this research.

**References:**

Alley, R. B., Shuman, C. A., Meese, D. A., Gow, A. J., Taylor, K. C., Cuffey, K. M., et al. (1997). Visual-stratigraphic dating of the GISP2 ice core: Basis, reproducibility, and application. *Journal of Geophysical Research: Oceans (1978–2012)*, *102*(C12), 26367–26381.

Barnes, P. R. F., Wolff, E. W., Mallard, D. C., & Mader, H. M. (2003). SEM studies of the morphology and chemistry of polar ice. *Microscopy Research and Technique*, *62*(1), 62–69. doi:10.1002/jemt.10385.

Cullen, D., & Baker, I. (2006). Observation of impurities in ice. *Microscopy Research and Technique*, *55*(3), 198–207.

Dedick AN, Hoffmann P. 1992. Chemical characterization of iron in atmospheric aerosols. Atmospheric Environment 26: 2545–2548.

Durand, G., Weiss, J., LIPENKOV, V., Barnola, J. M., Krinner, G., Parrenin, F., et al. (2006). Effect of impurities on grain growth in cold ice sheets. *Journal of Geophysical Research*, *111*(F1), F01015. doi:10.1029/2005JF000320

Faria, S. H., Freitag, J., & Kipfstuhl, S. (2010). Polar ice structure and the integrity of ice-core paleoclimate records. *Quaternary Science Reviews*, *29*(1-2), 338–351. doi:10.1016/j.quascirev.2009.10.016

Gow, A. J., Meese, D. A., Alley, R. B., Fitzpatrick, J. J., Anandakrishnan, S., Woods, G. A., & Elder, B. C. (1997). Physical and structural properties of the Greenland Ice Sheet Project 2 ice core: A review. *Journal of Geophysical Research: Oceans (1978–2012)*, *102*(C12), 26559–26575.

Obbard, R., Iliescu, D., Cullen, D., Chang, J., & Baker, I. (2003). SEM/EDS comparison of polar and seasonal temperate ice. *Microscopy Research and Technique*, *62*(1), 49–61. doi:10.1002/jemt.10381.

Mayewski, P. A., Sneed, S. B., Birkel, S. D., Kurbatov, A. V., & Maasch, K. A. (2013). Holocene warming marked by abrupt onset of longer summers and reduced storm frequency around Greenland. *Journal of Quaternary Science*, *29*(1), 99–104. doi:10.1002/jqs.2684

Mayewski PA, Meeker LD, Twickler MS, et al. 1997. Major features and forcing of high latitude northern hemisphere atmospheric circulation over the last 110,000 years. Journal of Geophysical Research 102: 26,345–26,366.

Meeker, L. D., & Mayewski, P. A. (2002). A 1400-year high-resolution record of atmospheric circulation over the North Atlantic and Asia. *The Holocene*, *12*(3), 257–266. doi:10.1191/0959683602hl542rp

Meeker LD, Mayewski PA, Twickler MS, et al. 1997. A 110,000-year history of change in continental biogenic emissions and related atmospheric circulation inferred from the Greenland Ice Sheet Project Two ice core. Journal of Geophysical Research 102: 26,489–26,504.

Mulvaney R., Wolff, E.W., Oates K. 1988. Sulphuric acid at grain bound- aries in Antarctic ice. Nature 331:247–249.

Reinhardt, H., Kriews, M., Miller, H., Ludke, C., Hoffmann, E. and Skole, J. (2003). Application of LA-ICP-MS in polar ice studies. Analytical Bioanalytical Chemistry 375, 1265–1275

Reinhardt, H., Kriews, M., Miller, H., Schrems, O., Ludke, C., Hoffmann, E. and Skole, J. 2001. Laser ablation inductively coupled plasma mass spectrometry: a new tool for trace element analysis in ice cores. Fresnius Journal of Analytical Chemistry 370, 629–636.

Rempel, A. W. (2002). Anomalous diffusion of multiple impurity species: Predicted implications for the ice core climate records. *Journal of Geophysical Research*, *107*(B12), 2330. doi:10.1029/2002JB001857.

Rempel, A. W., Waddington, E. D., Wettlaufer, J. S., & Worster, M. G. (2001). Possible displacement of the climate signal in ancient ice by premelting and anomalous diffusion. *Nature*, *411*(6837), 568–571.

Rempel, A. W., & Wettlaufer, J. S. (2003). Segregation, transport, and interaction of climate proxies in polycrystalline ice. *Canadian Journal of Physics*, *81*(1-2), 89–97. doi:10.1139/p02-118.

Sneed, Sharon B., Mayewski, Paul A., Sayre, W.G., Handley, Michael J., Kurbatov, Andrei V., Taylor, Kendrick C., Bohleber, Pascal, Wagenbach, Dietmar, Erhardt, Tobias, and Spaulding, Nicole E. (2015). New LA-ICP-MS cryocell and calibration technique for sub-millimeter analysis of ice cores. Journal of Glaciology. Volume 61, issue 226, 233-242.

Wolff, E. W. in Chemical Exchange Between the Atmosphere and Polar Ice (eds Wolff, E. W. & Bales, R. C.) 541 ± 560 (NATO ASI Series I, Vol. 43, Springer, Berlin, 1996).

Wolff, E.W., Mulvaney. R., Oates, K. 1988. The location of impurities in Antarctic ice. Ann Glaciol 11:194–197.